# Electromagnetic Scattering Laws in Weyl Systems

Ming Zhou[1], Lei Ying[1], Ling Lu[2], Lei Shi[3], Jian Zi[3], Zongfu Yu[1]

1. Department of Electrical and Computer Engineering, University of Wisconsin Madison, 53705, USA
2. Institute of Physics, Chinese Academy of Sciences, Beijing, China
3. Department of Physics, Fudan University, Shanghai, China

**Abstract**: Wavelength determines the length scale of the cross section when electromagnetic waves are scattered by an electrically small object. The cross section diverges for resonant scattering, and diminishes for non-resonant scattering, when wavelength approaches infinity. This scattering law explains the color of the sky as well as the strength of a mobile phone signal. We show that such wavelength scaling comes from free space's conical dispersion at zero frequency. Emerging Weyl systems, offering similar dispersion at non-zero frequencies, lead to new laws of electromagnetic scattering that allow cross sections to be decoupled from the wavelength limit. Diverging and diminishing cross sections can be realized at any target wavelength in a Weyl system, providing unprecedented ability to tailor the strength of wave-matter interactions for radio-frequency and optical applications.

Electromagnetic scattering is a fundamental process that occurs when waves in a continuum interact with an electrically small scatterer. Scattering is weak under non-resonant conditions; an example is Rayleigh scattering, which is responsible for the colors of the sky. Conversely, scattering becomes much stronger with resonant scatterers, which have an internal structure supporting localized standing waves, such as antennas, optical nanoresonators, and quantum dots. Resonant scatterers have wide application because resonance allows physically small scatterers to capture wave energy from a large area. As such, large electromagnetic cross sections, $\sigma$, are always desirable: a larger $\sigma$ value means (for example) stronger mobile phone signals[1] and higher absorption rates for solar cells[2].

The maximum cross section of resonant scattering is bounded by the fundamental limit of electrodynamics. One might be tempted to enlarge the scatterer to increase the cross section, but this strategy only works for non-resonant scattering, or with scatterers that are larger than the wavelength. In resonant scattering, physical size only affects spectral bandwidth, while the limit of cross section is determined by the resonant wavelength $\lambda$ as[3]:

$$\sigma_{max} = \frac{D}{\pi}\lambda^2 \qquad (1)$$

The directivity, $D$, describes the anisotropy of the scattering; $D = 1$ for isotropic scatterers. Equation 1 shows that an atom[4] can have a $\sigma_{max}$ similar to that of an optical antenna[5], despite the former's sub-nanometer size. This also means that optical scatterers cannot attain resonant cross sections as large as those of radio-frequency (RF) antennas, due to the smaller wavelengths involved.

Overcoming the limit just described has far-reaching implications for RF and optoelectronic applications. Many efforts have been devoted to realizing this goal, including the use of enhanced directivity $D$[6–12], degenerate resonances[13] and decreased dielectric constants[14] $\epsilon$. While these approaches exploit certain tradeoffs to slightly increase the prefactor in Eq. (1), the fundamental limit of $\lambda^2$ remains, which can be proven directly from Maxwell's equations without requiring specific scatterer details[5]. Until now, extremely large cross sections have only been obtained at long wavelengths near the DC frequency, as shown in Fig. 1a.

In this letter, we show that the free-space dispersion relation leads to the wavelength limit shown by Eq. (1). Specifically, the DC point, which is located at the apex of the conical dispersion relation, gives rise to the diverging cross sections at low frequencies. When the apex point is moved to another spectral location, as shown in Fig. 1b, the special scattering properties around the DC point move as well, resulting in diverging cross sections at high frequencies. Weyl points[15–23], the three-dimensional (3D) analog of Dirac points, have recently been shown to exhibit such conical dispersion relations. We derive the scattering laws in Weyl systems and show that their unique dispersion allows the cross section to be decoupled from the wavelength limit. This opens a new path to realizing strong wave-matter interactions, providing potential benefits to RF and optoelectronic devices that rely on resonant scattering.

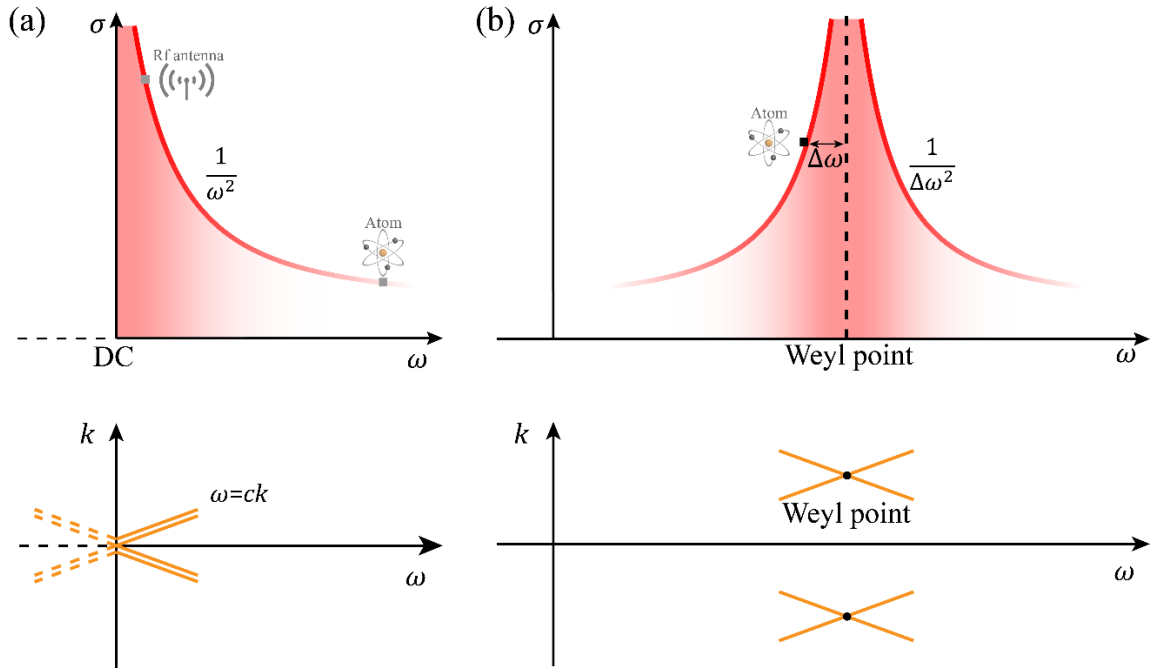

**Figure 1. The scale of cross section and its relation to dispersion**. (a) In free space, the resonant cross section scales according to $\sigma \sim \lambda^2$ or $\sim 1/\omega^2$. Large cross sections always favor low frequencies, and diverging cross sections are obtained around the DC point, which happens to be the apex of a conical dispersion. The cross section of a resonant transition in an atom is small ($\sim 10^{-12}$ m$^2$) because of the associated short wavelength ($\sim \mu$m). The cross section of an RF antenna is much larger ($\sim 10^{-4}$ m$^2$) due to a much longer wavelength ($\sim$cm). (b) By constructing new media with conical dispersions located away from the DC point, the associated resonant cross sections scale completely differently—diverging cross sections can be realized even at optical frequencies.

We start by considering the resonant cross section of a dipole antenna (Fig. 2a). Without losing generality, we only discuss the scattering cross section, assuming zero absorption. Similar conclusions can be drawn for the absorptive case, with the maximum absorption cross section $\frac{1}{4}$ that of the scattering cross section[24].

The dipole antenna is anisotropic due to its elongated shape, so the cross section $\sigma(\theta, \varphi)$ depends on the incident direction of the wave. At a normal direction, when $\theta = \pi/2$, it reaches its maximum value[5] of $\sigma_{max} = 3\lambda^2/2\pi$. Along the axial direction, when $\theta = 0$, the cross section vanishes. While it is straightforward to calculate the cross section of a dipole antenna, it is not immediately apparent why the cross section follows the $\lambda^2$ rule and diverges around the DC frequency. To illustrate the underlying physics, we will first demonstrate the general conservation law of resonant scattering.

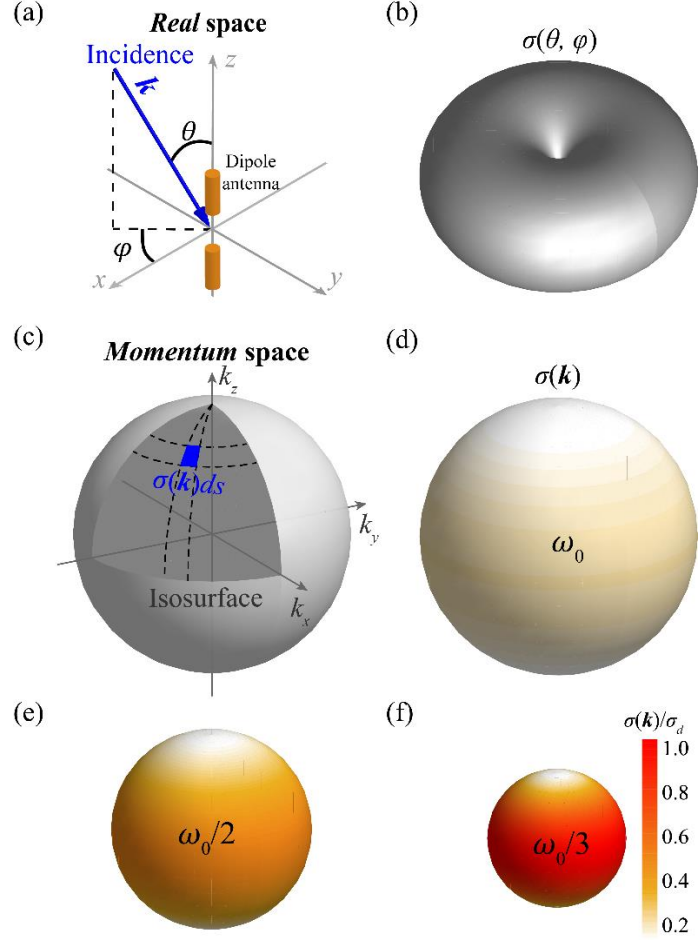

**Figure 2. Scaling law of resonant cross sections in momentum space.** (a) Schematic of a dipole antenna in free space. (b) Real-space representation of scattering cross section $\sigma(\theta,\varphi)$. (c) Mapping of the scattering cross section $\sigma(\boldsymbol{k})$ in momentum space. (d-f) Momentum-space representations of the scattering cross sections $\sigma(\boldsymbol{k})$ for different frequencies; the color intensity indicates the value of $\sigma(\boldsymbol{k})$, which is normalized by $\sigma_d = 27\lambda^2/2\pi$.

Figure 2b shows the real-space representation of $\sigma(\theta,\varphi)$ at the resonant frequency $\omega_0$. We can also represent this in momentum space, $\sigma(\boldsymbol{k})$, with $\boldsymbol{k}$ located on the isosurface defined by $\omega = \omega_0$. As shown in Fig. 2d, the isosurface is a sphere with $|\boldsymbol{k}| = \omega/c$. To visualize the momentum-space representation, $\sigma(\boldsymbol{k})$ is indicated by the color intensity on the isosurface in Fig. 2e. The resonant cross section satisfies the following conservation law:

$$\iint_{s:\,\omega(\boldsymbol{k})=\omega_0} \sigma(\boldsymbol{k})ds = 16\pi^2 \qquad (2)$$

The integration is performed on the frequency isosurface $\omega(\boldsymbol{k}) = \omega_0$. As shown in the Supplementary Information, Eq. 2 is generally true for resonant scattering. Our proof is based on quantum electrodynamics, so it applies to classical scatterers like antennas, as well as to quantum scatterers such as electronic transitions that absorb and emit light. More importantly, the continuum does not need to be free space; it can be anisotropic materials, or even photonic crystals, as long as a well-defined dispersion relation $\omega = \omega(\mathbf{k})$ exists.

Equation 2 dictates the scaling of $\sigma$ with respect to the scatterer's resonant frequency $\omega_0$. As $\omega_0$ decreases, the area of the isosurface shrinks. To conserve the value of the integration over a smaller isosurface, the cross section $\sigma(\mathbf{k})$ must increase accordingly. For example, Fig. 2e and 2f show the isosurfaces of dipole antennas with resonant frequencies at $\omega_0/2$ and $\omega_0/3$, respectively. The cross sections, indicated by color intensity, must increase proportionally to maintain a constant integration over these smaller isosurfaces. This can also be seen by the average cross section in the momentum space:

$$\bar{\sigma} \equiv \frac{\iint \sigma(\mathbf{k}) ds}{\iint ds} = \frac{16\pi^2}{S} \tag{3}$$

Here $S \equiv \iint ds$ is the area of the isosurface. When approaching the apex of the conical dispersion relation, the isosurface diminishes, i.e. $S \to 0$. As a result, the cross section diverges around the DC point (Fig. 1a).

Recent demonstrations of Weyl points in photonic crystals[16] show that the dispersion of a 3D continuum can exhibit conical dispersion at any designed frequency (Fig. 1b). The Hamiltonian for the continuum around the Weyl point is given by $\mathcal{H}(\mathbf{q}) = v_x q_x \sigma_x + v_y q_y \sigma_y + v_z q_z \sigma_z$, where $\sigma_{x,y,z}$ are Pauli matrices. The momentum $\mathbf{q}(q_x, q_y, q_z) = \mathbf{k} - \mathbf{k}_{Weyl}$ defines the distance to the Weyl point, and $\mathbf{q} = 0$ at the Weyl point. The linear dispersion described by this Hamiltonian produces an ellipsoidal isosurface that encloses the Weyl point. The isosurface shrinks to a point at the Weyl frequency $\omega_{Weyl}$. Scattering properties associated with the DC point are carried to high frequencies within Weyl media, resulting in exceptionally strong resonant scattering. As illustrated in Fig. 1b, the average cross section scales as $\bar{\sigma} \sim \frac{1}{(\omega - \omega_{Weyl})^2}$ around the Weyl point. This allows high frequency resonant scatterers, such as atoms and quantum dots, to attain large cross sections, potentially at a macroscopic scale.

Scattering by an object embedded in photonic crystals is much more complex than that in free space. The waves are highly dispersive, anisotropic, and non-uniform, and the scattering depends strongly on the scatterer's location and orientation within the photonic crystal. Remarkably, the conservation law as shown by Eq. (2) and (3) is independent of these variables.

We now demonstrate a specific example of resonant scattering in a Weyl photonic crystal[20]. Its structure consists of two gyroids, as shown in Fig. 3a. The red gyroid is defined by the equation $f(\mathbf{r}) > 1.1$, where $f(\mathbf{r}) = \sin(2\pi x/a)\cos(2\pi y/a) + \sin(2\pi y/a)\cos(2\pi z/a) + \sin(2\pi z/a)\cos(2\pi x/a)$ and $a$ is the lattice constant. The blue gyroid is the spatial inversion of the red one. The two gyroids are filled with a material with a dielectric constant of $\epsilon = 13$. To obtain Weyl points, we add four air spheres to the gyroids to break the inversion symmetry (Fig. 3a). These spheres are related by an $S_4(z)$ transformation. The resulting band structure has four isolated Weyl points at the same frequency of $\omega_{Weyl} = 0.5645\ (2\pi c/a)$. Fig. 3b illustrates conical dispersion in the $2k_z = -k_x - k_y$ plane.

The resonant scatterer is a quantum two-level system (TLS) embedded in the above photonic crystal. We numerically solve the scattering problem using quantum electrodynamics[25,26]. The Hamiltonian is $H = H_{pc} + H_{TLS} + H_I$. The first two terms, $H_{pc} = \sum_q \hbar\omega_q c_q^\dagger c_q$ and $H_{TLS} = \hbar\omega_0 a^\dagger a$, are the Hamiltonian of the photons and the TLS, respectively[27]. Here $\hbar$ is the reduced Planck constant, $a^\dagger$ and $a$ are the raising and lowering operators for the quantum dot, respectively, and $c_q^\dagger$ and $c_q$ are the bosonic creation and annihilation operators of the photons, respectively. The Lamb shift[27] is incorporated into the resonant frequency $\omega_0$. The third term, $H_I = i\hbar \sum_q g_q(c_q^\dagger a - c_q a^\dagger)$, is the interaction between the TLS and the radiation[27]. The coupling coefficient is $g_q = \boldsymbol{d} \cdot \widehat{\boldsymbol{E}}_q \sqrt{\omega_0/2\hbar\epsilon_0 L^3}$, where $\boldsymbol{d}$ is the dipole moment, $\epsilon_0$ is vacuum permittivity, $\widehat{\boldsymbol{E}}_q$ is the unit polarization vector of the photons, and $L^3$ is the quantization volume.

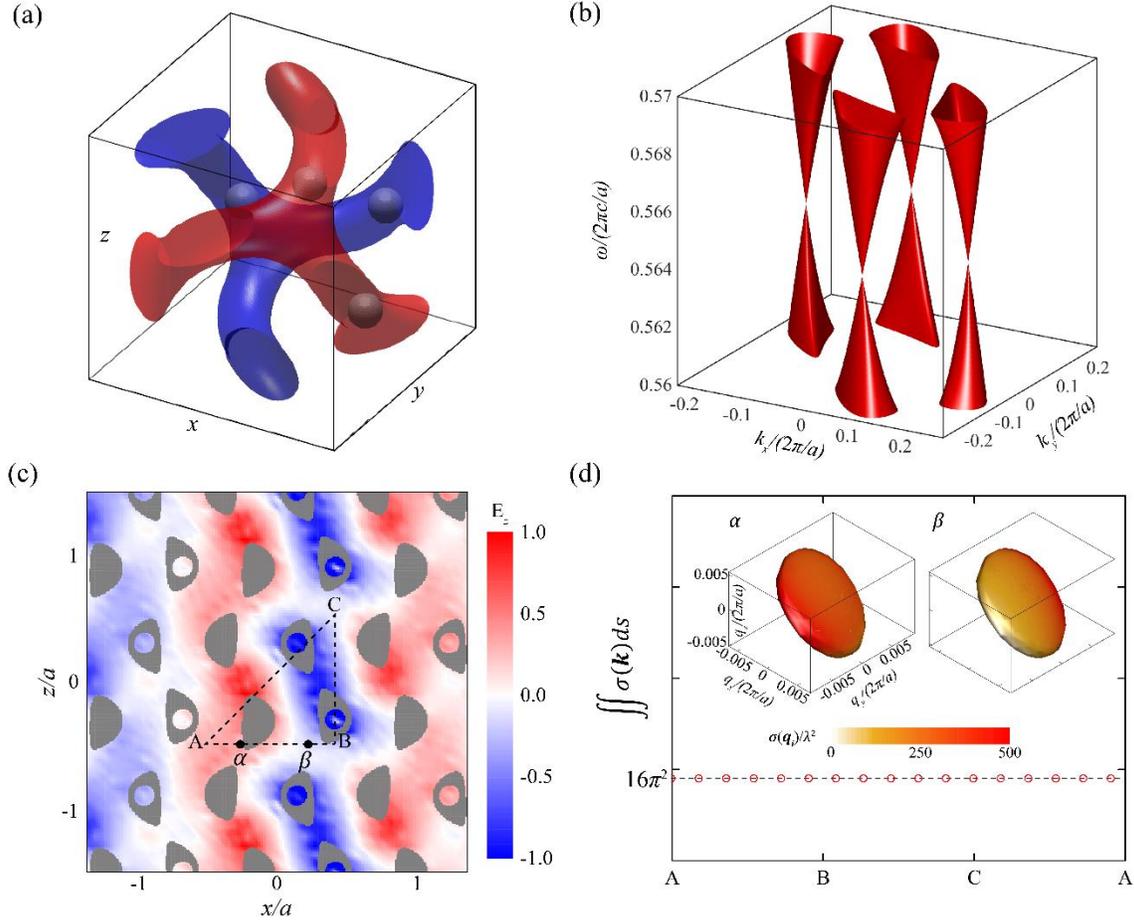

**Figure 3. Simultion of the quantum scattering of a two-level system embeded in Weyl point photonic crystals.** (a) The unit cell of the photonic crystal that supports Weyl points. Four air spheres with a radius of 0.07a are added to the gyroids to break the inversion symmetry. The positions of the air spheres are given by $(\frac{1}{4}, -\frac{1}{8}, \frac{1}{2})a$, $(\frac{1}{4}, \frac{1}{8}, 0)a$, $(\frac{5}{8}, 0, \frac{1}{4})a$ and $(\frac{3}{8}, \frac{1}{2}, \frac{1}{4})a$. (b) Band structure of the photonic crystal close to the Weyl

points. The band structure is plotted on the plane $2k_z = -k_x - k_y$. (c) The electric-field distribution of one eigenmode of the photonic crystal on the $x$-$z$ plane. (d) The resonant cross sections of the TLS for different locations. The integration in momentum space always leads to the same constant. The dipole moment of the TLS is in the $x$ direction. Positions A-B-C-A are also labeled in (c). Examples of $\sigma(\mathbf{q})$ at two different positions, $\alpha$ and $\beta$, are plotted on the isosurfaces as insets.

As an example, we consider a TLS with a transition frequency $\omega_0$ slightly below the Weyl frequency: $\omega_0 - \omega_{Weyl} = -0.0005\ (2\pi c/a)$. The calculated cross section $\sigma(\mathbf{q})$ (see Supplementary) is plotted on the isosurface as shown by inset-$\alpha$ of Fig. 3d. As expected, it strongly depends on the incident wavevector $\mathbf{q}$. In addition, $\sigma(\mathbf{q})$ varies greatly at different locations, as shown by comparing inset-$\alpha$ and inset-$\beta$ in Fig. 3d. Despite all these differences, when integrated over the isosurface, $\iint \sigma(\mathbf{q})ds$ always results in the same constant: $16\pi^2$. We perform the integration for a TLS at 20 different locations, all with the same constant, as shown in Fig. 3d.

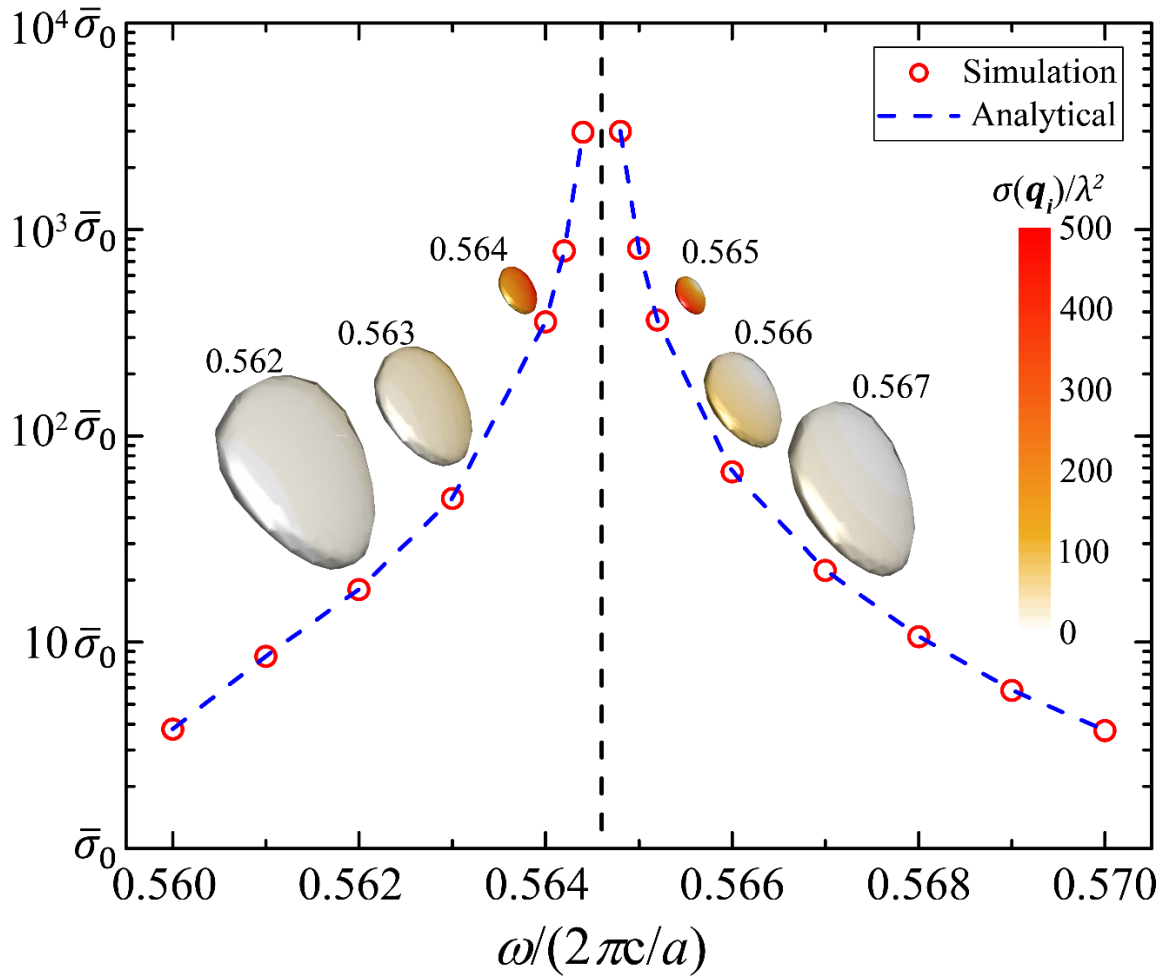

**Figure 4. Diverging resonant cross section is realized around the Weyl frequency.** The results from a direct quantum scattering simulation (red circles) agree well with the prediction based on the band structure (blue dashed line). The cross section is normalized by the average cross section in free space $\bar{\sigma}_0 = \lambda^2/\pi$. It scales as $\sigma \sim 1/\Delta\omega^2$ and diverges at the Weyl frequency. Isosurfaces have an ellipsoidal shape with its color indicating the value of the cross section; shrinking isosurface leads to increasing cross section.

As the TLS's transition frequency approaches the Weyl point, *i.e.*, $\omega \to \omega_{Weyl}$, the isosurface shrinks in size, as illustrated by the insets of Fig. 4. The conservation law leads to an increasing $\sigma(\boldsymbol{q})$, as shown by stronger colors. Near the Weyl frequency, the average cross sections $\bar{\sigma}$ is enhanced by three orders of magnitude compared to that in free space, eventually diverging at the Weyl point. The analytical prediction from Eq. 2 and the area of the isosurface agree very well with predictions from numerical simulation (circles in Fig. 4a).

*Rayleigh (non-resonant) scattering.* The non-resonant scattering of electrically small objects in free space follows the Rayleigh scattering law. In great contrast to resonant scattering, the cross section of Rayleigh scattering scales as $\sigma \sim \omega^4$, and diminishes at the DC frequency, as shown in Fig. 5a. This property is also closely related to the isosurface and can be carried to high frequency at the Weyl point (see proof in the Supplementary Information). While resonant cross section diverges, non-resonant cross section diminishes at Weyl points, as illustrated in Fig. 5b.

The total power of the scattered field in Rayleigh scattering can be shown as[28–30]:

$$P_0 = \omega^2 \iint_{S:\,\omega(\boldsymbol{k})=\omega} m(\boldsymbol{k}) ds \qquad (4)$$

Here the $m(\boldsymbol{k}) = \frac{\pi}{4\epsilon_0 v_{g,\boldsymbol{k}}} |\boldsymbol{a}_{\boldsymbol{k}}(\boldsymbol{r}_0) \cdot \boldsymbol{d}|^2$ coefficient is determined by the induced dipole moment $\boldsymbol{d}$, the field amplitude $\boldsymbol{a}_{\boldsymbol{k}}(\boldsymbol{r}_0)$, and the group velocity $v_{g,\boldsymbol{k}}$, which is in turn calculated from the eigenmode in the continuum at the position of the scatter $\boldsymbol{r}_0$. While $m(\boldsymbol{k})$ generally does not vary significantly with frequency, the integral $\iint_{S:\,\omega(\boldsymbol{k})=\omega_0} m(\boldsymbol{k}) ds$ is proportional to the area of the isosurface $S$. In free space, $S \sim \omega^2$. At the DC point, the isosurface shrinks to a point with $S = 0$, and the cross section of Rayleigh scattering is zero. Similarly, around the Weyl point, the area of isosurface $S \sim \Delta\omega^2 = (\omega - \omega_{Weyl})^2$. The cross section scales as $\sigma \sim \omega^2 \Delta\omega^2$, and is zero at the Weyl point (Fig. 5b).

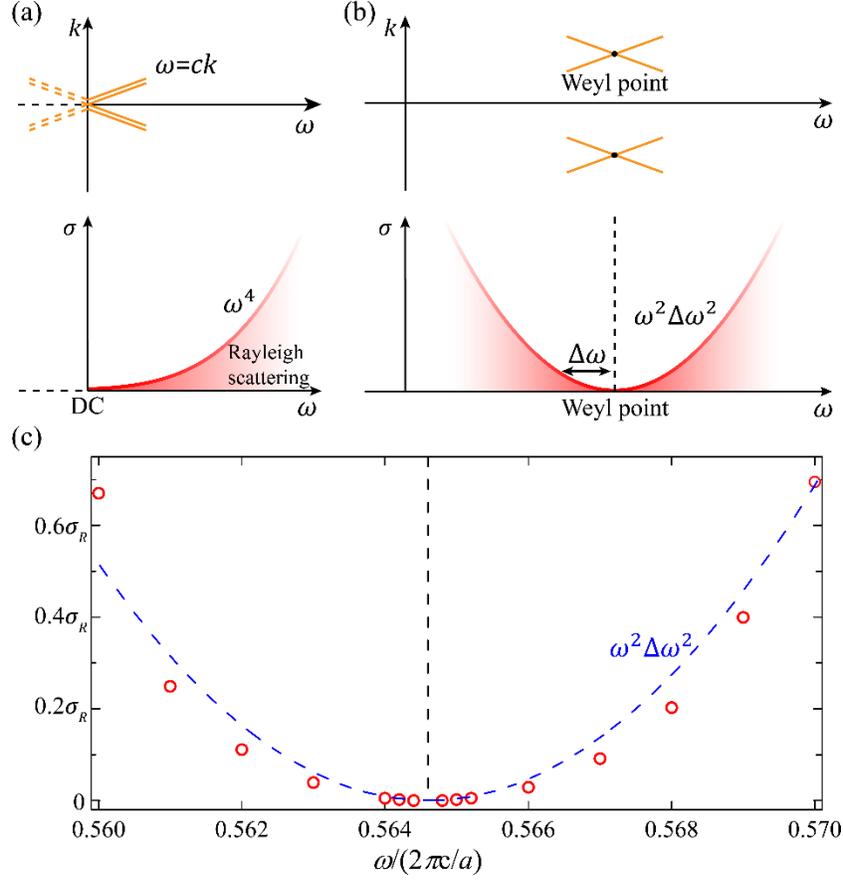

**Figure 5. Rayleigh scattering in free space and Weyl systems.** (a) In free space, the Rayleigh scattering cross section scales as $\sigma \sim \omega^4$ and vanishes at the DC frequency. (b) In Weyl systems, the Rayleigh scattering cross section scales as $\sigma \sim \omega^2 \Delta\omega^2$ and vanishes at the Weyl frequency. (c) Numerical simulation of Rayleigh scattering in a Weyl photonic crystal. The Rayleigh scatter is a dielectric sphere with a radius of $0.01a$ and a dielectric constant of 2. The numerical results (red circles) are normalized by the Rayleigh scattering cross section in free space $\sigma_R$, and fit well the predicated $\sigma \sim \omega^2 \Delta\omega^2$ scaling (blue dashed line). The differences at frequencies far away from the Weyl frequency result from the different shapes of isosurfaces (insets of Fig. 4).

To validate our theoretical prediction above, we numerically calculate the Rayleigh scattering cross section of a small dielectric sphere embedded in a Weyl photonic crystal. For consistency, we use the same Weyl photonic crystal we considered previously. The dielectric sphere has a radius of $0.01a$ and a dielectric constant of 2. We use MPB[31] to obtain the eigenmode of the Weyl photonic crystal, then numerically calculate the Rayleigh scattering cross section (see more details in the Supplementary Information). The calculated Rayleigh scattering cross sections around the Weyl point are obtained by averaging 100 scatterers at random locations within the photonic crystal, and are plotted in Fig. 5c as red circles. They are normalized by $\sigma_R$, the scattering cross section of the same scatter in free space. The blue dashed line indicates the scaling law $\sigma \sim \omega^2 \Delta\omega^2$. The calculated Rayleigh scattering cross section agrees with the theoretical prediction well, with a vanishing cross section observed at the Weyl point. For high frequency devices, such as integrated waveguides and laser cavities, Rayleigh scattering caused by interface roughness degrades performance and increases noise. The combination of suppressed Rayleigh scattering and enhanced resonant scattering could make Weyl media attractive for these optoelectronic applications.

Finally, the conservation law of resonant scattering can also be extended to lower-dimensional space. In two-dimensional photonic crystals, diverging cross sections can be realized at Dirac points; an example is provided in the Supplementary Information. Two-dimensional crystals are easier to use for experimental demonstration than three-dimensional Weyl media. To further generalize the findings in this paper, we may not necessarily need conical dispersion. Quadratic dispersion found around the band edges of photonic crystals, also provides shrinking isosurfaces. However, this is less useful in practice because the zero group velocity at the band edge makes it difficult to obtain propagating waves[32] in the presence of disorders. In addition, coupling into such media is difficult due to the large impedance mismatch.

In conclusion, we show that a Weyl point, as the apex of a conical dispersion relation, has similar scattering properties as those around the DC point. Exceptionally strong resonant scattering is accompanied by diminishing non-resonant scattering. It was believed that the cross section was dictated by the length scale of the wavelength, but here, since Weyl points can be realized at any frequency, we can effectively decouple the two. Weyl media opens up a new direction for tailoring wave-matter interaction with unprecedented flexibility.

Acknowledgement